\documentclass{optica-article}
\graphicspath{ {images/} }
\journal{opticajournal} 
\articletype{Research Article}
\usepackage{lineno}
\usepackage{amsmath, bm}

\begin{document}
\title{B-spline freeform surface tailoring for prescribed irradiance based on differentiable ray-tracing}

\author{Haoqiang Wang,\authormark{1} Zihan Zang,\authormark{1,2} Yunpeng Xu,\authormark{1} Yanjun Han,\authormark{1,3,4} Hongtao Li\authormark{1} and Yi Luo\authormark{1,2,3,4,5}}

\address{\authormark{1} Beijing National Research Centre for Information Science and Technology, Department of Electronic Engineering, Tsinghua University, Beijing 10084, China\\
	\authormark{2} Tsinghua-Berkeley Shenzhen Institute (TBSI), Tsinghua University, Shenzhen 518055, China\\
	\authormark{3}Center for Flexible Electronics Technology, Tsinghua University, Beijing 100084, China\\
	\authormark{4}Flexible Intelligent Optoelectronic Device and Technology Center, Institute of Flexible Electronics Technology of THU, Zhejiang, Jiaxing 304006, China\\}

\email{\authormark{5}luoy@tsinghua.edu.cn}


\begin{abstract*} 
A universal and flexible design method for freeform surface that can modulate the distribution of an zero-étendue source to an arbitrary irradiance distribution is a significant challenge in the field of non-imaging optics. 
Current design methods typically formulate the problem as a partial differential equation and solve it through sophisticated numerical methods, especially for off-axis situations.
However, most of the current methods are unsuitable for directly solving multi-freeform surface or hybrid design problems that contains both freeform and spherical surfaces.
To address these challenges, we propose the B-spline surface tailoring method, based on a differentiable ray-tracing algorithm. 
Our method features a computationally efficient B-spline model and a two-step optimization strategy based on optimal transport mapping. 
This allows for rapid, iterative adjustments to the surface shape based on deviations between the simulated and target distributions while ensuring a smooth resulting surface shape.
In experiments, the proposed approach performs well in both paraxial and off-axis situations, and exhibits superior flexibility when applied to hybrid design case.
\end{abstract*}

\section{Introduction}
\begin{figure}[htb]
    \centering
    \includegraphics[width=0.5\textwidth]{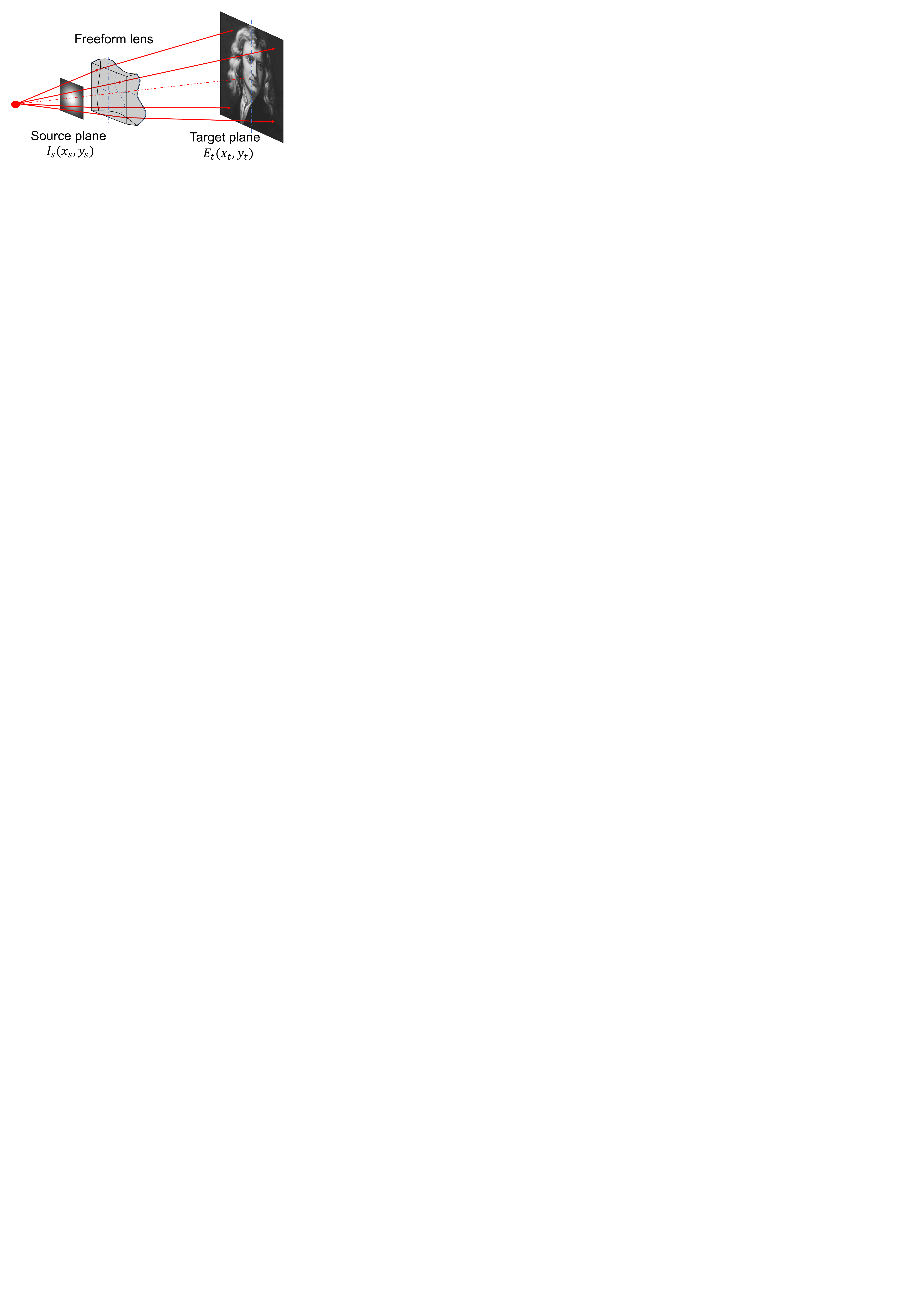}
    \caption{Schematic diagram of the non-imaging optical system produce prescribed irradaince distribution form a zero-etendue source.}
    \label{fig:system}
\end{figure}

One of the fundamental yet difficult problems in the non-imaging optics regime is designing freeform optical surfaces that can produce a specified irradiance distribution from a given light source.
A schematic diagram of this non-imaging optical system is shown in Fig.\ref{fig:system}.
The light emitted from the light source is refracted by the freeform surface, so that the intensity distribution of the source $I_s(x_s, y_s)$ is reshaped to a prescribed irradiance $E_t(x_t, y_t)$ on the target plane.

To derive the shape of the freeform lens, three critical constraints always need to be taken into account simultaneously.
Firstly, smoothness constraints of all surfaces are required to be considered to ensure they are machinable.
Secondly, Snell's law, which describes the rules for the propagation of light passing through freeform surfaces.
Thirdly, energy conservation between the source and target plane also needs to be ensured, which is formulated as a partial differential equation:
\begin{equation}
    \label{eq:mapping}
    I_s(x_s, y_s) = E_s(\phi (x_s, y_s)det(\nabla \phi(x_s, y_s))) 
\end{equation}
where $\phi: (x_s,y_s)\rightarrow(x_t,y_t)$ is a map between source and target plane.

The ray-mapping method\cite{constuct, Raymapping-2010, mapping16, review} is the most commonly used method for designing freeform surfaces, and it contains two steps.
Firstly, the method solves the energy conservation equation (Eq.\ref{eq:mapping}) to obtain ray-mapping $\phi: (x_s,y_s)\rightarrow(x_t,y_t)$.
Secondly, by combining the smoothness constraints and Snell's law, the shape of the freeform surface usually can be constructed point-by-point through numerical integration from a predefined initial point \cite{constuct}.
However, the ray-mapping that satisfies the energy conservation is not unique, and the greatest challenge of the method is to obtain an integrable ray-mapping.
The $L^2$ optimal transport mapping (OT-map) \cite{raymap-2017, Feng-OT} refers to the mapping that satisfies the energy conservation and minimizes the $L^2$ distance between the source and target coordinates.
In addition, there are many efficient numerical solvers \cite{ OT-idm,lsq, bfm} available for the OT-map and it has been well studied.
However, the OT-map is integrable only in paraxial approximation\cite{limitation}.

To address the limitations of the OT-map method, several modifications have been proposed.
For instance, one approach is a measure preserving transformation proposed in \cite{off-axis}, which alters the initial OT-map iteratively by integrating a Hamiltonian flow.
The goal of this alteration is to maintain the target distribution while changing the mapping to satisfy the smoothing constraint.
Another approach proposed by Wei et al. \cite{Wei-least-square} involves modifying the iterative process of the least-squares solver\cite{lsq} for OT-map by adding a medium step to make the mapping integrable.
Feng et al. \cite{Feng-IWT} proposed an iterative wavefront tailoring (IWT) method, which considers the smoothness constraints of the wavefront instead of the surface itself and derives an second order partial differential equation about the wavefront. 
This equation can be iteratively solved to obtain an integrable ray-mapping.
Although these methods allow the ray-mapping method to achieve complex irradiance distributions under off-axis situations, the additional constraints or solving steps that need to be introduced make the numerical process more complicated than the original OT-map method.

Differing from ray-mapping method, the Monge-Ampère (MA) method \cite{MA-reflect, MA-ol-2013} directly combines the above three critical constraints to establish a second-order partial differential equations of MA type for surface $S(x,y,f(x,y))$. 
However, while the MA method can produce complex irradiance distributions, deriving the MA equation can be complicated and time-consuming. 
Additionally, any changes in the system configuration will require the MA equation to be re-derived, which can be a significant challenge in practice.

In summary, the current method for freeform surface design uses partial differential equations to formulate the problem and solve it numerically.
The obtained freeform surfaces are usually represented as a finite number of sampled points and then fitted with a NURBS (Non-Uniform Rational B-Spline) model to create a machinable 3D model.
Moreover, these methods are typically limited to single-surface conditions and cannot be directly applied to multi-surface design problems without additional constrains, such as optical path length need to be considered to constrct multi-surface design \cite{2surf-feng, MA-double}.

In contrast to the methods in the field of non-imaging optics, the direct optimization method is commonly used for designing spherical or aspherical surfaces in imaging applications\cite{fischer_tadic-galeb_yoder_2008}.
This method formulates the surface using explicit expressions with some variable parameters and tailors the surface shape by directly updating the surface parameters according to the results of ray-tracing.
In recent decades, advancements in parallel computing hardware have led to significant improvements in the speed of ray-tracing.
Furthermore, the development of automatic differentiation techniques\cite{Autodiff} has made it easier to obtain the gradient between ray-tracing results and surface parameters, resulting in the emergence of a new type of ray-tracing algorithm called differentiable ray-tracing (DiffRT)\cite{Difflen,diffrt2,dO-wang,Nie:23,dekoning2023gradient}.
With modern gradient-based optimizers, the design of a imaging system with aspherical lenses  using DiffRT has been demonstrated by Sun et al. \cite{Difflen}. 
Wang et al. \cite{dO-wang} have optimized memory usage by not considering the iterative process of solving intersections between the ray and surface during automatic differentiation.
Nie et al. \cite{Nie:23} have implemented the XYPolynomial freeform surface model for designing reflective and refractive imaging optical systems.
However, only a few attempts have been made in the design of free-form surfaces for generating prescribed irradiances.
Wang et al. \cite{dO-wang} presented an example of rendering caustic images based on B-spline surface, but the results were significantly worse than traditional methods and it did not explain how to calculate the intersection points between rays and B-spline surfaces. 
Dekoning et al. \cite{dekoning2023gradient} also used a B-spline model, but had to approximate the B-spline surface with a triangular mesh to calculate the ray-surface intersection points, and the results also heavily relied on the initialization of parameters. 

In our work, we propose the B-spline surface tailoring (BST) method, which incorporates a computationally efficient, differentiable ray-tracing algorithm with a well-designed B-spline surface model.
In order to address the issue of strong reliance on initial shape of the freeform surfaces, the BST method present a two-step optimization strategy. 
The first steps derives an excellent inital shape based on OT-map and the second step tailors the shape according to the difference of the target and simulated irradaince distributions directly.
This method enables a efficient and accurate solution of complex irradiance distribution generation problems, as demonstrated through design examples presented in this study.
Section two provides detailed description on the direct optimization method based on differentiable ray-tracing, the proposed B-spline model, and the two-step optimization strategy.
In the following section, we present design examples for creating complex irradiance distributions under both paraxial and off-axis conditions.
We also demonstrate the capability of the proposed method for hybrid design problems that contains both spherical and freeform surfaces.

\section{Design Method}
\label{sec:2}
\subsection{ General direct optimization method based on differentiable ray-tracing }

\begin{figure}[htb]
    \centering
    \includegraphics[width=0.9\textwidth]{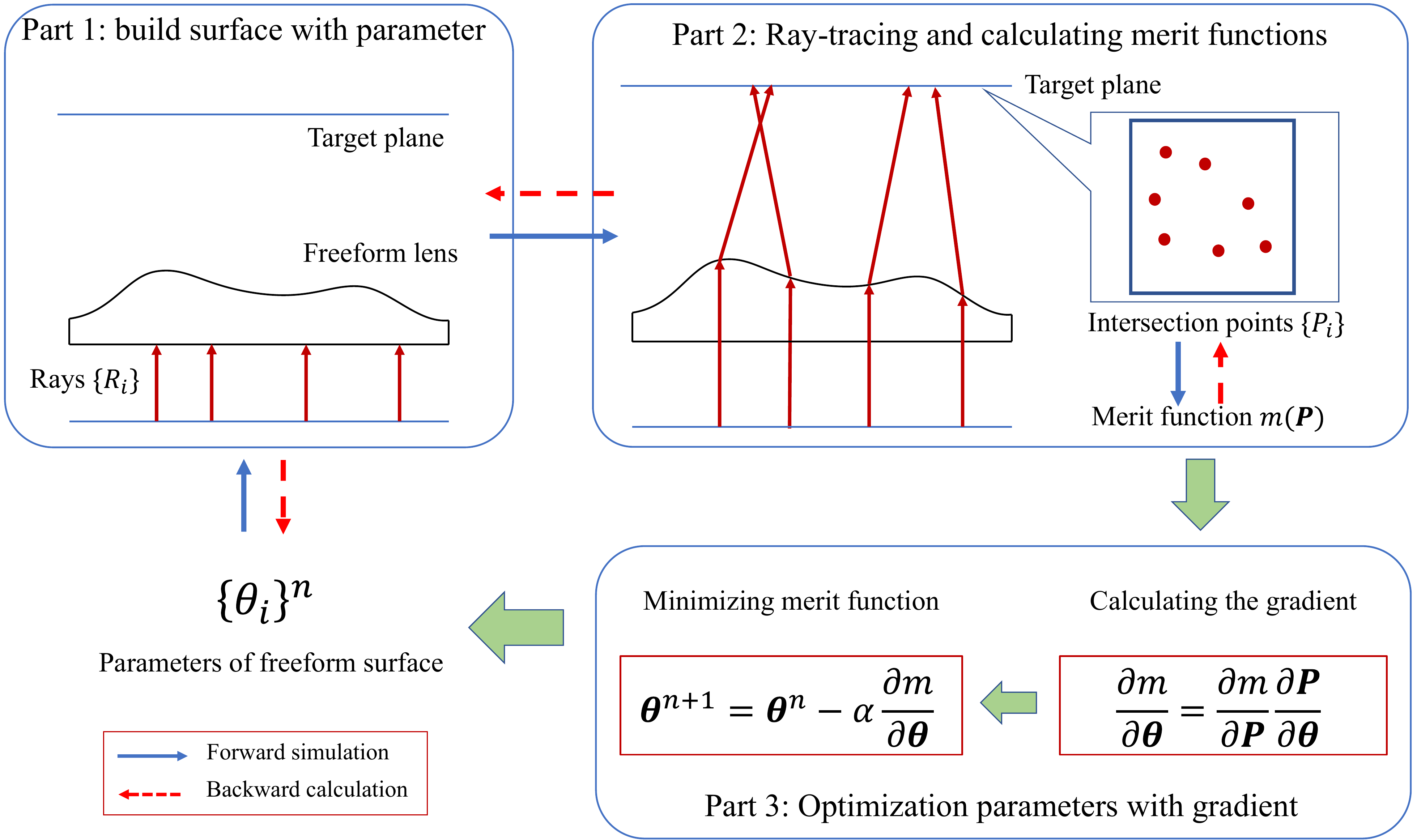}
    \caption{ Three parts of the general direct optimization method based on differentiable ray-tracing algorithm. }
    \label{fig:diffRT}
\end{figure}

In general, the procedure of the direct optimization method based on differentiable ray-tracing is illustrated in Fig.\ref{fig:diffRT} and contains three critical parts.
Firstly, suitable surface parameters $\left\{\theta_i\right\}$ and mathematical model are chosen to establish the structure of the optical system.
The position of light source and the target plane also need to be determined in accordance with the application requirements.

In the second part, rays $\left\{R_i\right\}$ emitted from the source plane are simulated to propagate and refract through the optical system, and then intersect with the target plane to obtain the intersection points $\left\{P_i(\bm{\theta})\right\}$.
The merit function, which evaluates the performance of the optical system from the intersection points, is also calculated in the second part.
Notably, the simulation results obtained in this part are consistent with the traditional Monte-Carlo ray-tracing algorithm. The core difference is that the simulation algorithm in this part is implemented using modern differentiable programming tools such as PyTorch\cite{pytorch}.

In the third part, based on the results of differentiable ray-tracing in part two, the gradient between the merit function and the parameters of the surface can be automatically derived.
Thereafter, gradient-based non-convex optimizer can be employed to update surface parameters to minimize the merit function.

It is important to note that for the implementation of the differentiable ray-tracing algorithm in freeform surface design, a general explicit mathematical model for freeform surfaces is required.
The surface model needs to satisfy the global smoothness of the surface while allowing for convenient local modifications.
Additionally, it is crucial to choose a suitable merit function that aligns with the physical nature of the system.

\subsection{ General and proposed B-spline models } 

To meet these requirements for surface model, the B-spline model is considered as the best choice. It has been widely used in computer-aided design and 3D reconstruction \cite{nurbs_3d}.
B-spline is short for basis spline, and it refers to a type of piece-wise polynomial function.
A typical B-spline surface can be formulated as follows \cite{nurbs}:
\begin{equation}
    \label{eq:surface}
    S(t_x, t_y) = \sum_i^m{\sum_j^m{ B_{i,n}(t_x)B_{j,n}(t_y) P_{i,j} }}
\end{equation}
where $P_{i,j}=(p_x, p_y, p_z)$ is the coordinates of the control points, $m$ is the number of control points of one axis (there are totally $m^2$ control points), $B_{i,n}(t)$ means the $i-th$ basis function of order $n$, which is a piece-wise polynomial function of degree $n-1$ with a parameter $t$.
The basis function of B-spline are typically defined and calculated by recursion \cite{nurbs}: 
\begin{equation}
    \label{eq:basis}
    B_{i,{k+1}}(t) = \omega_{i,k}(t)B_{i,k}(t) + [1-\omega_{i,k}(t)]B_{i+1, k}(t)
\end{equation}
where
\begin{equation}
    \label{eq:basis2}
    \omega_{i,k}(t)=
    \begin{cases}
        \frac{t-t_i}{t_{i+k}-t_i}, \quad &t_{i+k} \neq t_i, \\
        0,\quad & otherwise.
    \end{cases}
\end{equation}
and
\begin{equation}
    \label{eq:basis3}
    B_{i,1}(t)=
    \begin{cases}
        1,\quad &t_i \leq t < t_{i+1}, \\
        0,\quad & otherwise.
    \end{cases}
\end{equation}
The $\left\{t_0, t_1, ..., t_{m+n-1}\right\}$ in Eq.\ref{eq:basis} is known as knot vector and the knots $t_i$ are sorted into non-decreasing order, which are the position where the pieces of polynomial meet.

In addition, the normal vectors of B-spline surface are required in simulation and they can be derived by the derivatives of surface:
\begin{equation}
    \begin{aligned}
        \vec{N}(t_x, t_y) &= \vec{T_x}(t_x,t_y) \times \vec{T_y}(t_x,t_y) \\
           &= \sum_i^m{\sum_j^m{B^{'}_{i,n}(t_x)B_{j,n}(t_y)P_{i,j} }}\\
           &\times \sum_i^m{\sum_j^m{ B_{i,n}(t_x)B^{'}_{j,n}(t_y)P_{i,j} }}
    \end{aligned}
\end{equation}

In general, there are two main difficulties in implementing ray-tracing algorithm based on general B-spline surface.
Firstly, the recursive form of the surface expression, although suitable for programming implementation, is inefficient in numerical calculations. Deriving a general analytic expression is difficult due to the flexibility of control points and knots.
Secondly, it is also difficult to directly calculate the intersection between the equation of the ray and parametric equations of freeform surfaces, which requires to further obtain the relationship between parameter $t$ and real coordinates.
To overcome these issues, control points and knots are carefully defined, an analytic expression that can be efficiently calculated are further derived.

\begin{figure}[!htb]
    \centering
    \includegraphics[width=\textwidth]{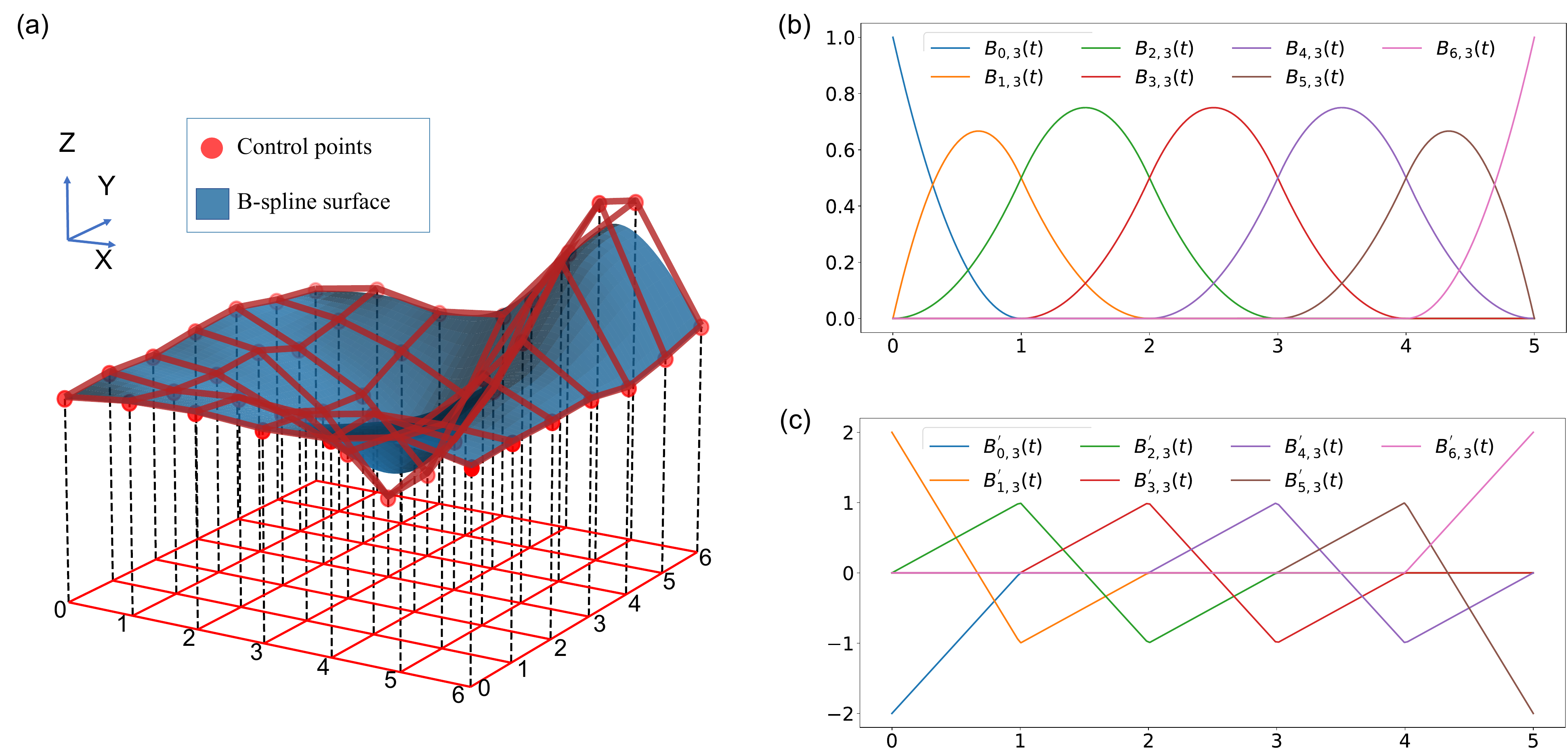}
    \caption{ (a) Proposed B-spline surface model with uniformly distributed control points. (b),(c) Visualization results of basis functions and derivative functions of the proposed B-spline model of order 3, with 7 control points. }
    \label{fig:b_spline}
\end{figure}

First, the knots vector of the B-spline surface are fixed uniformly, where the difference between knots is 1. For instance, for a B-spline surface of order $n$ with $m^2$ control points, the knot vector of one axis is:
\begin{equation}
    \left\{  \underbrace{0,...,0}_n,1,2,...,m-n-1,m-n, \underbrace{m-n+1,...,m-n+1}_n  \right\}
\end{equation}

In the second step, as shown in Fig. \ref{fig:b_spline}(a), the projection of the control points onto the $X-Y$ plane forms a fixed uniform mesh grid on the interval of $[0,m-1]\times[0,m-1]$, and only the z-coordinates of the control points are the variable parameters of the proposed B-spline surface model. 
In practice, the $X-Y$ coordinates are linearly transformed to match the real region of the surface while maintaining uniformity.

By introducing the above two constraints on the B-spline model of order 3, convenient analytic expression of this B-spline surface can be derived.
The basis function of B-spline model of order 3 can be write as:
\begin{equation}
    B_{i+2,3}(t)=
    \begin{cases}
        \frac{1}{2}(t-i)^2,\quad & i \leq t < i+1 \\
        -(t-i-1)^2 + (t-i-1) + \frac{1}{2}, \quad & i+1 \leq t < i+2 \\
        \frac{1}{2}(i+3-t)^2, \quad & i+2 \leq t < i+3 \\
        0 \quad & otherwise
    \end{cases}
\end{equation}
where $i\in[0,m-5]$. There are also four corner cases need to be mentioned:

\begin{equation}
    B_{0,3}(t)=
    \begin{cases}
        (1-t)^2,\quad & 0\leq t < 1 \\
        0 \quad & otherwise
    \end{cases}
\end{equation} 
\begin{equation}
    B_{1,3}(t)=
    \begin{cases}
        -\frac{3}{2}t^2+ 2t,\quad & 0\leq t < 1 \\
        \frac{1}{2}(2-t)^2,\quad & 1\leq t < 2 \\
        0 \quad & otherwise
    \end{cases}
\end{equation} 
\begin{equation}
    B_{m-1,3}(t)=
    \begin{cases}
        (m-2-t)^2,\quad & m-3\leq t < m-2 \\
        0 \quad & otherwise
    \end{cases}
\end{equation} 
\begin{equation}
    B_{m-2,3}(t)=
    \begin{cases}
        \frac{1}{2}(m-3-t)^2,\quad & m-4\leq t < m-3 \\
        -\frac{3}{2}(m-2-t)^2+ 2(m-2-t),\quad & m-3\leq t < m-2 \\
        0 \quad & otherwise
    \end{cases}
\end{equation}

The derivative of the basis function can also be easily obtained. Fig.\ref{fig:b_spline} (b,c) shows the visualization results of basis functions and them derivative when number of control points is 7.
Moreover, with the fixed control points and knots, an analytic expression for the relationship between the parameter $t$ and the real coordinates can be obtained :
\begin{equation}
    t(x)=
    \begin{cases}
        2-\sqrt{(4-2x)},\quad & 0\leq x < 3/2 \\
        x-\frac{1}{2},\quad & 3/2\leq x < m-5/2 \\
        m-4+\sqrt{4-2(m-1-x)},\quad & m-5/2\leq x < m-1 \\
    \end{cases}
\end{equation}
where the coordinate $x$ is normalized to  $[0,m-1]$ by $x = (m-1) \times (x_{real}-x_{min}) / (x_{max}-x_{min})$.
Using these analytic expression, B-spline surfaces of order 3 can be easily expressed and programmed to implement the corresponding algorithms for the intersection and reflection of rays.

\subsection{ Procedure of B-spline surface tailoring }

\begin{figure}[htb]
    \centering
    \includegraphics[width=\textwidth]{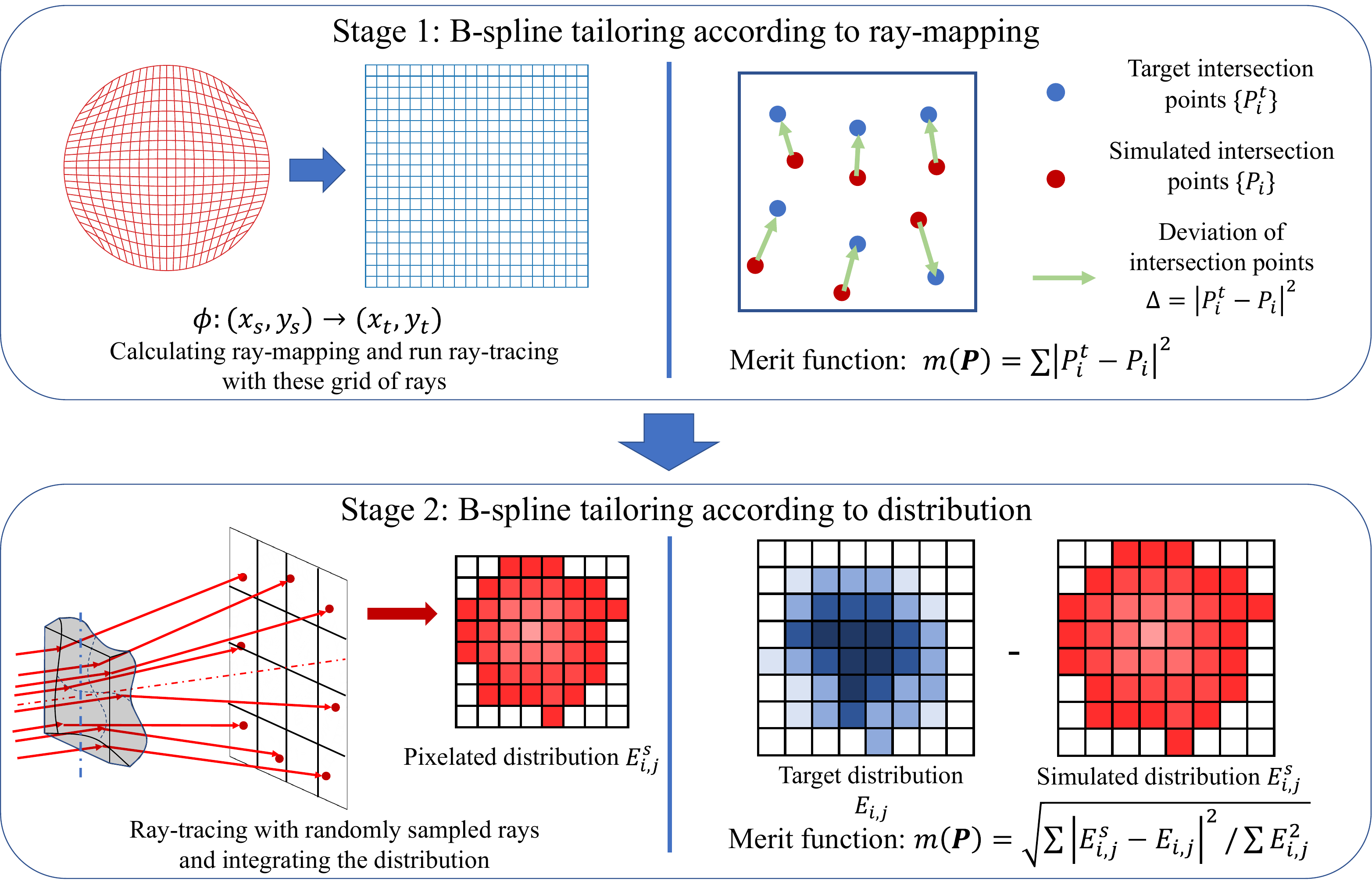}
    \caption{ Two stages of proposed B-spline surface tailoring method. In the first stage, the optimal transport mapping is derived and the shape of the freeform surface is tailored to match the mapping according to the deviations between the desired and simulated intersection points. In the second stage, using the shape from the first stage as an initial shape, the freeform surface is simulated with randomly sampled rays and tailored according to the difference between the desired and simulated irradiance distributions directly. }
    \label{fig:two_stage}
\end{figure}

After obtaining the B-spline model that can be efficiently calculated, in theory, the shape of the surface can be optimized by choosing the difference between the prescribed irradaince distribution and the actual simulated irradaince distribution as the merit function.
However, this first-effort method is prone to falling into local optima, and although the resulting surface is smooth, there will be many regions of large curvature on the surface, which is not conducive to be actual machinied.

In order to obtain a suitable initial surface shape and improve efficiency of the design of freeform surfaces.
The design process is divided into two stages, each using a different merit function.
As shown in Fig. \ref{fig:two_stage}, the first stage involves deriving a great initial shape based on the optimal transport mapping (OT-map).
The OT-map $\phi$ between the distribution of the source and target planes is calculated and two sets of corresponding grid points $\left\{S_i\right\}$ and $\left\{ P^t_i \right\}$ on the source and target plane are obtained firstly.
Rays $\left\{R_i\right\}$ emitting through $\left\{S_i\right\}$ are desired to be redirected by freeform surface and intersect with the target plane on $\left\{ P^t_i \right\}$.
These rays are used in the differentiable ray-tracing simulation process, and the merit function is defined as the $L^2$ norm between the real simulated intersection points $\left\{P_i(\bm{\theta})\right\}$ and the desired intersection points $\left\{ P^t_i \right\}$ derived by ray-mapping, which can be expressed as follows:

\begin{equation}
m(\mathbf{P}(\bm{\theta})) = \sum_i{|P^t_i - P_i|^2}
\end{equation}
After optimization, the merit function is minimized to obtain the best smooth freeform surface that matching the OT-map.

In the second stage, this surface derived in the first stage is used as the initial state.
A large number of randomly sampled rays are used in the simulation, and the merit function is chosen as the relative root mean square difference (RRMSD) between the normalized prescribed and simulated irradiance distributions:
\begin{equation}
m(\mathbf{P}(\bm{\theta})) = \sqrt{\sum_i^w\sum_j^w{|E^s_{i,j} - E_{i,j}|^2}/\sum_i^w\sum_j^w{E_{i,j}^2}},
\end{equation}
where $E_{i,j}$ and $E^{s}_{i,j}$ are the pixelated target and simulated irradiance distributions, respectively, and the simulated irradiance distribution $E^{s}_{i,j}$ is derived by numerically integrating the intersection points $\left\{P_i(\bm{\theta})\right\}$.
It is important to emphasize that this numerical integration process also needs to be differentiable, and this paper adopts the same strategy as in \cite{CongLi_PHD}, using a linear function to approximate the irradiance contributed by each intersection points.
During optimization process, the merit function is reduced and the simulated distribution approaches the prescribed distribution after a few optimization steps.
\section{Design example}
\subsection{Algorithm implementation}
To implement the proposed B-spline tailoring method, we developed a differential ray-tracing algorithm with proposed B-spline model using PyTorch \cite{pytorch}, an open-source machine learning framework that supports automatic differentiation and GPU acceleration.

It is worth emphasizing that, as shown in Fig.\ref{fig:b_spline}(b,c), for any value of parameters $t$, only the three adjacent basis functions are non-zero, and their derivatives have the same properties.
Therefore, instead of summing the products of all basis functions and control points coordinates according to Eq.\ref{eq:surface}, only these nine non-zero basis functions need to be compute to obtain the surface points.
This significantly reduces the computation time and memory usage, making the proposed method more efficient and scalable.

In the implementation process, the basis functions and their derivatives are stored as sparse matrices, and optimized sparse matrix algorithms\cite{pytorch} are used to accelerate the ray-tracing algorithm.
By utilizing sparse matrices in the simulation, the memory consumption is only dependent on the number of rays, rather than the number of control points used in the surface model.
This enables the use of more control points in the design, providing greater design freedom, without increasing the simulation time and memory consumption.

Moreover, in the first stage of B-spline surface tailoring, the OT-map is obtained through the back-and-forth method \cite{bfm}, which is a solver for the $L^2$ optimal transport problem with nearly linear time complexity.
In addition, the bellowing design example are run on a machine with Ryzen5 5600G CPU and NVIDIA RTXA4000 GPU with 16 GB GRAM.

\subsection{ Example of spherical and freeform surface hybrid design with ideal point source }

\begin{figure}[htb]
    \centering
    \includegraphics[width=\textwidth]{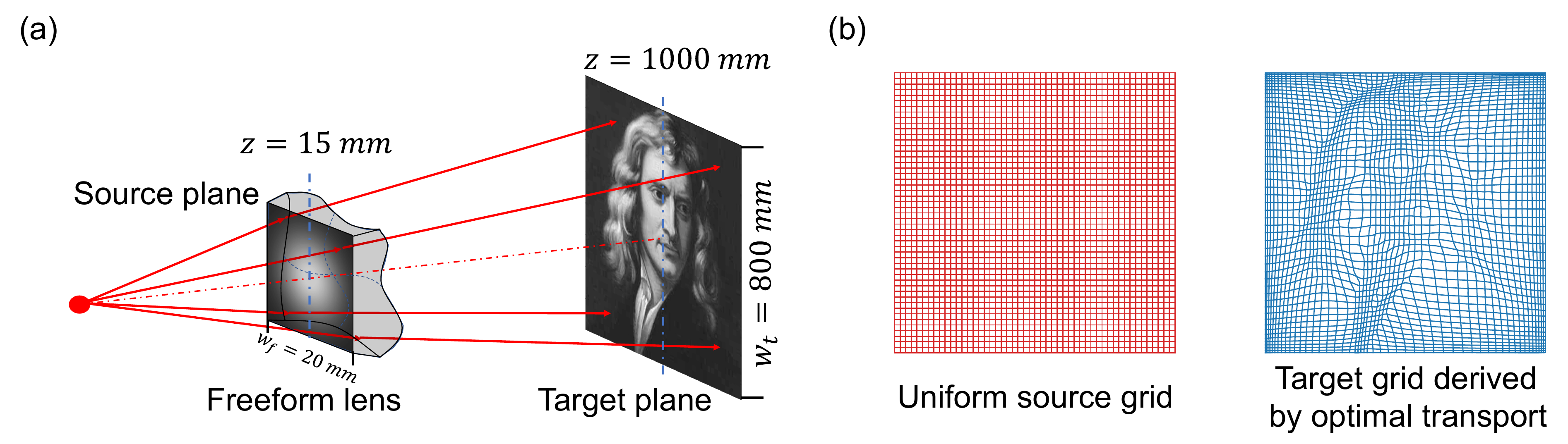}
    \caption{ (a) Schematic diagram of freeform optical system for the first design example with an ideal point source. (b) Optimal transport mapping between source and target irradiance distributions. }
    \label{fig:Example-1}
\end{figure}

We first focus on generating complex image pattern with an ideal point source.
The schematic of the optical system with a freeform lens is shown in Fig.\ref{fig:Example-1}(a), where the ideal point source is located at the origin $(0,0,0)$ and emits light of wavelength $650nm$ that produces a circular Gaussian beam with a $1/e^2$ divergence angle of $30^{\circ}$.
The light from this source is refracted by a freeform surface lens and finally intersects the target plane located at $z=1000mm$, forming the portrait of Isaac Newton in the region $[-400,400]\times[-400,400]$, with a resolution of $512\times 512$.

In this case, the material of the freeform lens is PMMA, which has a refractive index of 1.488 for the light source of $650nm$ , and its dimensions are $20mm \times 20mm$.
Furthermore, the front surface being a spherical surface and the back surface being a freeform surface, located at $z=15$ and $z=20$, respectively.

It is important to note that both front and back surfaces contribute to the refraction of the rays emitted by the light source and require simultaneous optimization. 
Although this front spherical surface has a small effect, it requires a redesign of the corresponding surface construction method for ray mapping method other than re-derived the differential equation for MA equation method.
However, for the B-spline surface tailoring method proposed in this paper, this change does not affect the optimization process.

Additionally, in this design example, both the spherical and freeform surfaces are represented by their heights relative to the initial plane.
The model for the spherical surface is given by \cite{fischer_tadic-galeb_yoder_2008}:
\begin{equation}
S(x,y) = \frac{C(x^2+y^2)}{1+\sqrt{1-C^2(x^2+y^2)}}
\end{equation}
where $C$ represents the curvature of the spherical surface.
The back freeform surface is modeled using the B-spline model proposed in this paper. 
During the surface tailoring process, the optimization step size is small, resulting in little deviation from the initial surface shape. 
In fact, no additional constraints were introduced by selecting appropriate initial positions to avoid intersection between the front and back surfaces.

\begin{figure}[!htb]
    \centering
    \includegraphics[width=\textwidth]{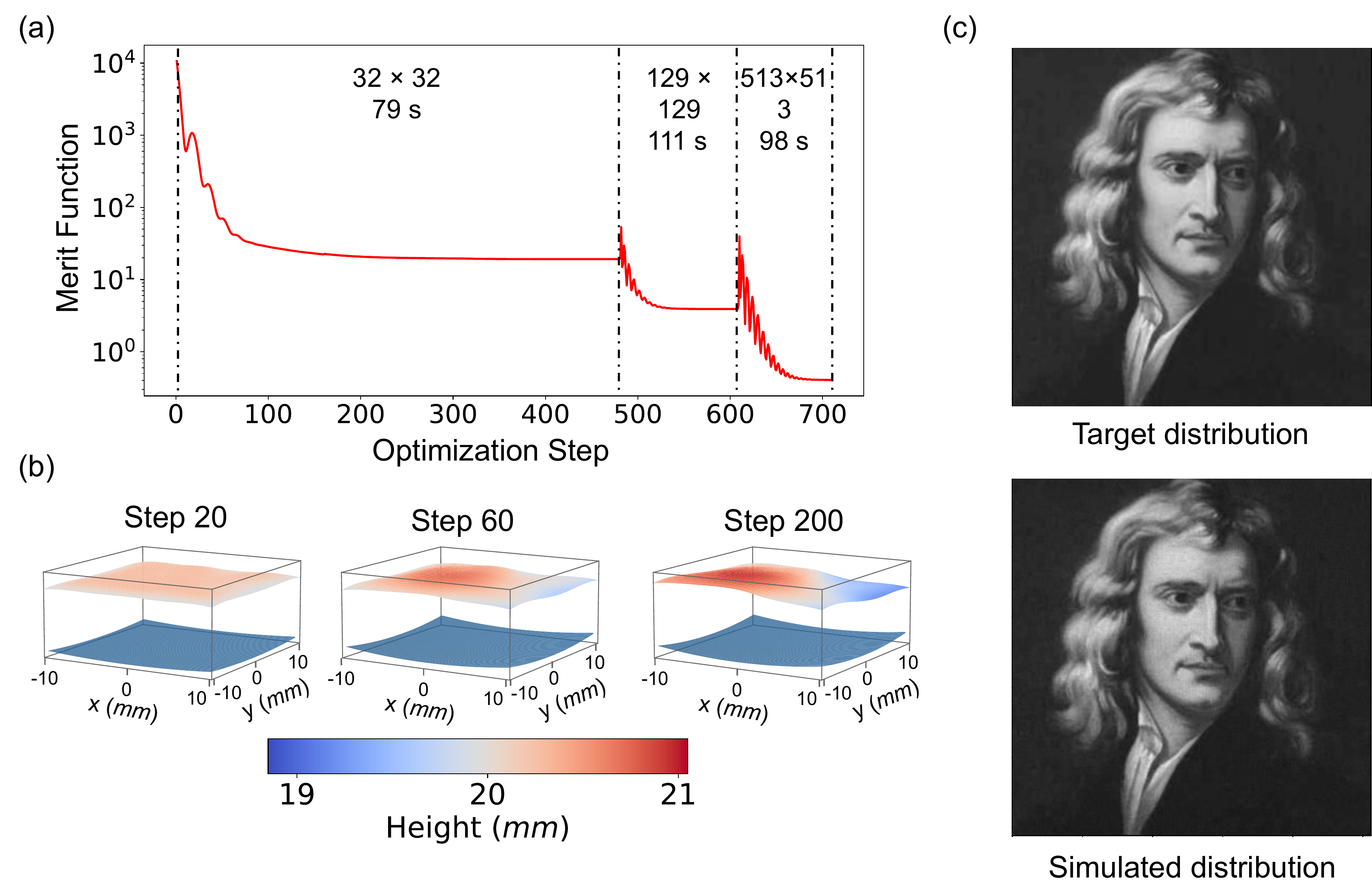}
    \caption{ The first stage of the B-spline surface tailoring method for the first design example (a) The deviation of the target and simulated intersection points decrease during optimization. (b) The shape of B-spline surface changed rapidly to minimizing the merit function. (c) Ray-tracing testing result with random sampled rays, and the RRMSD between simulated and target distributions is 12.3\% }
    \label{fig:Example-1-stage-1}
\end{figure}

In the first stage of the design process, the optimal transport mapping based on the expected irradiance distribution from the light source to the target plane is obtained.
The result is shown in Fig.\ref{fig:Example-1}(b), with the left subfigure showing the uniform grid points on the source plane, and the right subfigure showing the corresponding grid points on the target plane obtained through the optimal transport mapping.
The numerical solution accuracy is $1001 \times 1001$.

Next, a set of rays emitted from the origin and passing through the uniform grid points on the virtual source plane at $z=15mm$ were constructed for differentiable ray-tracing simulation.
The expected intersection points of these rays with the target plane are the grid points on the target plane mentioned above.

In the first stage of the surface tailoring process, the front and back surface of freeform lens was all flat surfaces initially.
The back freeform surface used $513 \times 513$ control points, and a multi-scale optimization strategy was adopted to make the surface modification process more robust.
First, the freeform surface with $33 \times 33$ control points was optimized.
After completing one round of optimization, the number of control points was increased to $129 \times 129$ and then further increased to $513 \times 513$.
The Adam\cite{adam} optimizer was used for the above tailoring process, with learning rates of 0.02, 0.004, and 0.008 used for the three scales, respectively. 
As for the front spherical surface, only the curvature is variable and the learning rate is chosen to 0.001.

Fig.\ref{fig:Example-1-stage-1} (a) shows the changes in the merit function during the optimization process.
The deviation between the simulated the target intersection points decreases rapidly, and the shape of the freeform surface and spherical surface changed quickly while remaining smooth, as shown in Fig.\ref{fig:Example-1-stage-1} (b).
Fig.\ref{fig:Example-1-stage-1} (c) shows the irradaince distribution obtained by ray-tracing simulation with randomly sampled rays after completing the first stage of the surface tailoring.
The RRMSD between the target and simulated irradiance distribution is 12.17\%, which is small enough and doesn't need to further second stage optimization.

\subsection{ Example of Off-axis illumination design with collimated source }

\begin{figure}[htb]
    \centering
    \includegraphics[width=\textwidth]{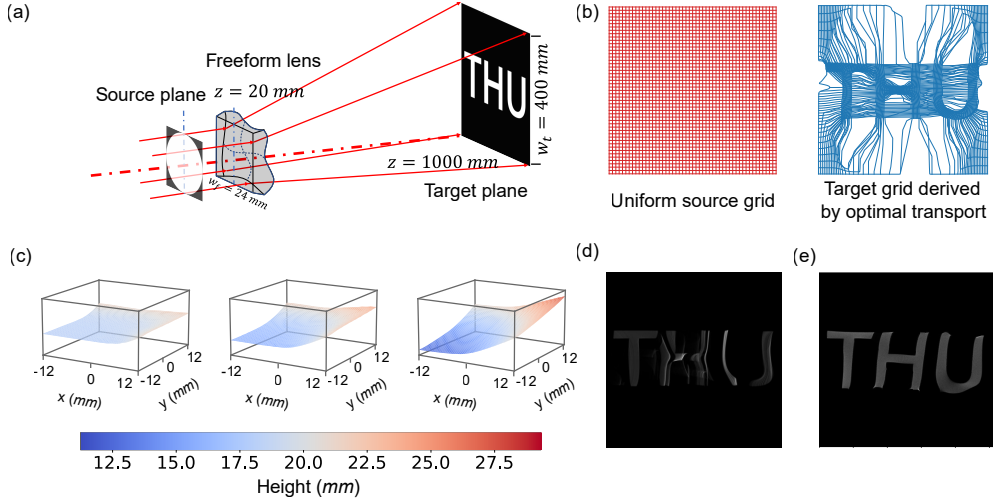}
    \caption{ (a) System setup of the off-axis example for collimated source. (b) Optimal transport mapping between source and target irradiance distributions. (c) The shape of B-spline surface changed rapidly to matching the rays derived by optimal transport mapping. (d) Simulated irradaince distribution using surface constructed by point-by-point integrating method. (e) Simulated irradaince distribution using surface constructed with the first stage optimization of the B-spline surface tailoring method. }
    \label{fig:THU-setup}
\end{figure}

In the second example, a off-axis illumination problem with collimated source is considered.
The system setup of this example is shown in Fig.\ref{fig:THU-setup}.(a). 
A uniform collimated source with radius $12mm$ is used and the wavelength is also $650nm$.
The freeform lens is a "plane-concave" lens and only the back surface is freeform surface, which is located at $z=20mm$ initially.
The dimensions of the freeform lens is $24mm \times 24mm$. 
The target distribution is the images of uniform alphabets "THU" with a resolution of $512\times 512$, which is located at $Z=1000mm$ plane and its region is $[0,400] \times [0,400]$. 

It should be emphasized that the target irradiance distribution in this example is discontinuous (with regions of zero energy), which makes the optimal transport mapping theoretically contain singular values. 
Using commonly used solvers for continuous optimal transport problems makes it difficult to achieve high-quality mapping solutions. 
Constructing surfaces based on such mappings is even more challenging.
Fig.\ref{fig:THU-setup}(b) shows the optimal transport mapping calculated with back-and-forth method between the uniform source distribution and the target irradiance distribution, and it is evident that the quality of this mapping is not very good.
When using the traditional point-by-point integration method to construct the surface, the ray-tracing simulation results shows significant distortion and non-continuous areas, shown in Fig.\ref{fig:THU-setup}(d).
It is due to the combined effect of off-axis situation and poor quality of the optimal transport mapping.

Fig.\ref{fig:THU-setup}(c) shows the optimization process of the freeform surface in the first stage of the proposed B-spline surface tailoring method.
The shape of the surface gradually changes from the initial plane to the expected tilted surface while maintaining smoothness.
Fig.\ref{fig:THU-setup}(e) shows the result of ray-tracing using the surface obtained after the first-stage optimization.
The irradiance distribution does not exhibit any discontinuity, only weak distortion in the output pattern caused by off-axis situation.
It is evident that the initial shape obtained through B-spline tailoring method is better than the surface shape obtained through point-by-point integration.

\begin{figure}[htb]
    \centering
    \includegraphics[width=\textwidth]{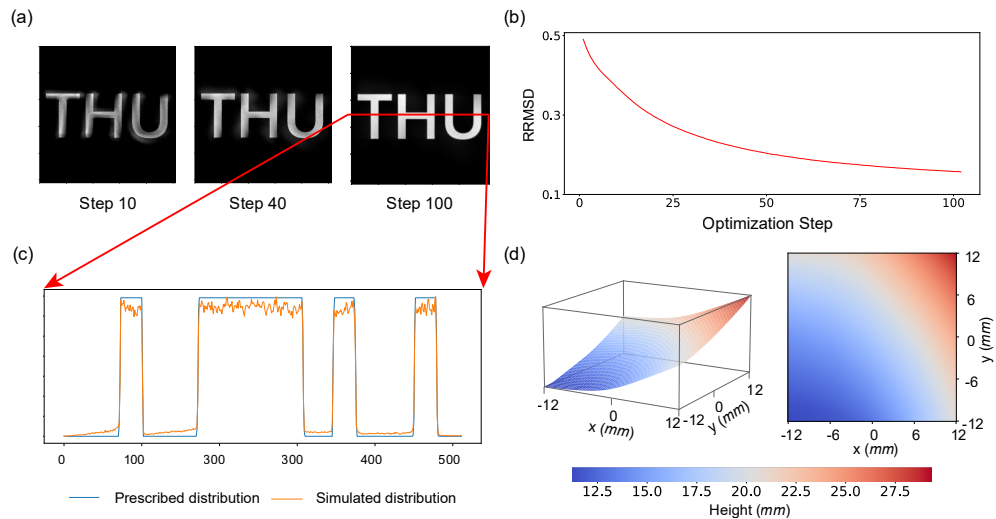}
    \caption{ The second stage of the B-spline surface tailoring method for the second design example. (a) Simulated irradaince distributions during the second stage of the B-spline surface tailoring (b) The RRMSD between the target and simulated irradaince distributions decrease during optimization. (c) Target and simulated irradaince distributions along the line $y=200mm$ (d) 3D visualization result and the height map of the desired freeform surface after tailoring. }
    \label{fig:THU_stage-2}
\end{figure}

In the second stage of the B-spline surface tailoring, random sampled rays is used in ray-tracing simulation and obtain the corresponding output irradiance distribution.
The shape of the surface will be directly modified based on the deviation between the simulated and the target irradiance distributions. Fig.\ref{fig:THU_stage-2} (a) shows the change in the simulated irradiance distributions during the optimization process.
It can be seen that the degree of distortion in the output pattern gradually decreases and becomes more uniform.
The RRMSD between the simulated and target illuminance distributions rapidly decreases from 48\% to 12.54\%, as shown in Fig.\ref{fig:THU_stage-2}(b).
Fig.\ref{fig:THU_stage-2} (c) further illustrates the distribution of the final simulated irradiance and the target irradiance along the line $y=200mm$ on the target plane.
Obviously, these two distributions are very close, although there is still some background energy between the letters.
Fig.\ref{fig:THU_stage-2}(d) shows the final shape of the freeform surface.

\section{Conclusion}
In these examples, the capability of our B-spline surface tailoring method for generating arbitrary complex distribution under both paraxial and off-axis situations are demonstrated.
Unlike the common used methods mentioned in Sec.1, which formulate freeform surface design problem in terms of differential equations, the proposed optimization approach allows for effective surface shape optimization based on the deviations between
the simulated and the target distributions while ensuring the smoothness of surfaces.
For ideal point and collimated sources, no additional algorithmic adjustments are required, except the defining the sampled rays emitted from the source. 
This method also worked well for the case of freeform lens contains front spherical surface and back freeform surface, which shows great potential to become a universal freeform surface design method.

In addition, this method could be used for extended light source theoretically if we could find an effective method to sampling the light emitted by a extended light source.
More complex optical field control, which control the irradaince distribution and the wavefront in the same time, also could be achieved by designing a suitable merit function.
These two application scenarios will be considered in our future works.
Further improving the speed of the ray-tracing algorithm and improving the stability of the optimization process will also be an important direction of our future work.

\begin{backmatter}
\bmsection{Funding} National Natural Science Foundation of China (61875104).

\bmsection{Disclosures} The authors declare no conflicts of interest.

\bmsection{Data availability} Data underlying the results presented in this paper are not publicly available at this time but may be obtained from the authors upon reasonable request.
\end{backmatter}

\bibliography{diff_rt}

\begin{thebibliography}{10}
\newcommand{\enquote}[1]{``#1''}

\bibitem{constuct}
Y.~Luo, Z.~Feng, Y.~Han, and H.~Li, \enquote{{Design of compact and smooth
  free-form optical system with uniform illuminance for LED source},}
  {\protect\JournalTitle{Optics Express}} \textbf{18}, 9055 (2010).

\bibitem{Raymapping-2010}
F.~R. Fournier, W.~J. Cassarly, and J.~P. Rolland, \enquote{{Freeform reflector
  design using integrable maps},} {\protect\JournalTitle{Optics Express}}
  \textbf{18}, 5295--5304 (2010).

\bibitem{mapping16}
C.~B{\"{o}}sel and H.~Gross, \enquote{{Ray mapping approach for the efficient
  design of continuous freeform surfaces},} {\protect\JournalTitle{Optics
  Express}} \textbf{24}, 14271 (2016).

\bibitem{review}
R.~Wu, Z.~Feng, Z.~Zheng, R.~Liang, P.~Ben{\'{i}}tez, J.~C. Mi{\~{n}}ano, and
  F.~Duerr, \enquote{{Design of Freeform Illumination Optics},}
  {\protect\JournalTitle{Laser \& Photonics Reviews}} \textbf{12}, 1700310
  (2018).

\bibitem{raymap-2017}
K.~Desnijder, P.~Hanselaer, and Y.~Meuret, \enquote{{Flexible design method for
  freeform lenses with an arbitrary lens contour},}
  {\protect\JournalTitle{Optics Letters}} \textbf{42}, 5238 (2017).

\bibitem{Feng-OT}
Z.~Feng, B.~D. Froese, and R.~Liang, \enquote{{Freeform illumination optics
  construction following an optimal transport map},}
  {\protect\JournalTitle{Applied Optics}} \textbf{55}, 4301 (2016).

\bibitem{OT-idm}
J.-D. Benamou, B.~D. Froese, and A.~M. Oberman, \enquote{Numerical solution of
  the optimal transportation problem using the monge{\textendash}amp{\`{e}}re
  equation,} {\protect\JournalTitle{Journal of Computational Physics}}
  \textbf{260}, 107--126 (2014).

\bibitem{lsq}
C.~R. Prins, R.~Beltman, J.~H.~M. ten Thije~Boonkkamp, W.~L. IJzerman, and
  T.~W. Tukker, \enquote{A least-squares method for optimal transport using the
  monge--ampère equation,} {\protect\JournalTitle{SIAM Journal on Scientific
  Computing}} \textbf{37}, B937--B961 (2015).

\bibitem{bfm}
M.~Jacobs and F.~L{\'{e}}ger, \enquote{A fast approach to optimal transport:
  the back-and-forth method,} {\protect\JournalTitle{Numerische Mathematik}}
  \textbf{146}, 513--544 (2020).

\bibitem{limitation}
A.~Bruneton, A.~B{\"{a}}uerle, R.~Wester, J.~Stollenwerk, and P.~Loosen,
  \enquote{{Limitations of the ray mapping approach in freeform optics
  design},} {\protect\JournalTitle{Optics Letters}} \textbf{38}, 1945 (2013).

\bibitem{off-axis}
K.~Desnijder, P.~Hanselaer, and Y.~Meuret, \enquote{{Ray mapping method for
  off-axis and non-paraxial freeform illumination lens design},}
  {\protect\JournalTitle{Optics Letters}} \textbf{44}, 771 (2019).

\bibitem{Wei-least-square}
S.~Wei, Z.~Zhu, Z.~Fan, and D.~Ma, \enquote{{Least-squares ray mapping method
  for freeform illumination optics design},} {\protect\JournalTitle{Optics
  Express}} \textbf{28}, 3811 (2020).

\bibitem{Feng-IWT}
Z.~Feng, D.~Cheng, and Y.~Wang, \enquote{{Iterative wavefront tailoring to
  simplify freeform optical design for prescribed irradiance},}
  {\protect\JournalTitle{Optics Letters}} \textbf{44}, 2274 (2019).

\bibitem{MA-reflect}
C.~R. Prins, J.~H. M. T.~T. Boonkkamp, J.~van Roosmalen, W.~L. Jzerman, and
  T.~W. Tukker, \enquote{A monge--amp{\`{e}}re-solver for free-form reflector
  design,} {\protect\JournalTitle{{SIAM} Journal on Scientific Computing}}
  \textbf{36}, B640--B660 (2014).

\bibitem{MA-ol-2013}
R.~Wu, L.~Xu, P.~Liu, Y.~Zhang, Z.~Zheng, H.~Li, and X.~Liu, \enquote{Freeform
  illumination design: a nonlinear boundary problem for the elliptic
  monge{\textendash}amp{\'{e}}re equation,} {\protect\JournalTitle{Optics
  Letters}} \textbf{38}, 229 (2013).

\bibitem{2surf-feng}
Z.~Feng, L.~Huang, G.~Jin, and M.~Gong, \enquote{Designing double freeform
  optical surfaces for controlling both irradiance and wavefront,}
  {\protect\JournalTitle{Optics Express}} \textbf{21}, 28693 (2013).

\bibitem{MA-double}
Y.~Zhang, R.~Wu, P.~Liu, Z.~Zheng, H.~Li, and X.~Liu, \enquote{Double freeform
  surfaces design for laser beam shaping with monge{\textendash}amp{\`{e}}re
  equation method,} {\protect\JournalTitle{Optics Communications}}
  \textbf{331}, 297--305 (2014).

\bibitem{fischer_tadic-galeb_yoder_2008}
R.~E. Fischer, B.~Tadic-Galeb, and P.~R. Yoder, \emph{Optical system design}
  (SPIE Press, 2008).

\bibitem{Autodiff}
A.~Paszke, S.~Gross, S.~Chintala, G.~Chanan, E.~Yang, Z.~DeVito, Z.~Lin,
  A.~Desmaison, L.~Antiga, and A.~Lerer, \enquote{Automatic differentiation in
  pytorch,} in \emph{NIPS-W,}  (2017).

\bibitem{Difflen}
Q.~Sun, C.~Wang, F.~Qiang, D.~Xiong, and H.~Wolfgang, \enquote{End-to-end
  complex lens design with differentiable ray tracing,}
  {\protect\JournalTitle{ACM Transactions on Graphics (TOG)}} \textbf{40}
  (2021).

\bibitem{diffrt2}
Z.~Li, Q.~Hou, Z.~Wang, F.~Tan, J.~Liu, and W.~Zhang, \enquote{End-to-end
  learned single lens design using fast differentiable ray tracing,}
  {\protect\JournalTitle{Optics Letters}} \textbf{46}, 5453 (2021).

\bibitem{dO-wang}
C.~Wang, N.~Chen, and W.~Heidrich, \enquote{{dO: A differentiable engine for
  Deep Lens design of computational imaging systems},}
  {\protect\JournalTitle{IEEE Transactions on Computational Imaging}}
  \textbf{8}, 905--916 (2022).

\bibitem{Nie:23}
Y.~Nie, J.~Zhang, R.~Su, and H.~Ottevaere, \enquote{Freeform optical system
  design with differentiable three-dimensional ray tracing and unsupervised
  learning,} {\protect\JournalTitle{Opt. Express}} \textbf{31}, 7450--7465
  (2023).

\bibitem{dekoning2023gradient}
B.~de~Koning, A.~Heemels, A.~Adam, and M.~Möller, \enquote{Gradient
  descent-based freeform optics design using algorithmic differentiable
  non-sequential ray tracing,}  (2023).

\bibitem{pytorch}
A.~Paszke, S.~Gross, F.~Massa, A.~Lerer, J.~Bradbury, G.~Chanan, T.~Killeen,
  Z.~Lin, N.~Gimelshein, L.~Antiga, A.~Desmaison, A.~Kopf, E.~Yang, Z.~DeVito,
  M.~Raison, A.~Tejani, S.~Chilamkurthy, B.~Steiner, L.~Fang, J.~Bai, and
  S.~Chintala, \enquote{Pytorch: An imperative style, high-performance deep
  learning library,} in \emph{Advances in Neural Information Processing Systems
  32,}  (Curran Associates, Inc., 2019), pp. 8024--8035.

\bibitem{nurbs_3d}
E.~Dimas and D.~Briassoulis, \enquote{3d geometric modelling based on {NURBS}:
  a review,} {\protect\JournalTitle{Advances in Engineering Software}}
  \textbf{30}, 741--751 (1999).

\bibitem{nurbs}
L.~Piegl and W.~Tiller, \emph{The NURBS Book} (Springer-Verlag, New York, NY,
  USA, 1996), 2nd ed.

\bibitem{CongLi_PHD}
C.~Wang, \enquote{Computational wavefront sensing: Theory, practice, and
  applications,} Ph.D. thesis, King Abdullah University of Science and
  Technology, Thuwal, Kingdom of Saudi Arabia (2021).

\bibitem{adam}
D.~P. Kingma and J.~Ba, \enquote{Adam: A method for stochastic optimization,}
  (2014).

\end{thebibliography}

\end{document}